\documentclass[conference]{article}

	\usepackage[margin=1.2in]{geometry}
	\usepackage[printonlyused,smaller]{acronym}
	\usepackage{dsfont}
	\usepackage{amsfonts}
	\usepackage{amssymb}
	\usepackage{tikz-cd}
	\usepackage{tikz}
	\usetikzlibrary{arrows,automata}
	\usepackage{float}
	\usepackage{mathtools}
	\usepackage{wrapfig}
	\usepackage{enumitem}
	\usepackage[bookmarksnumbered=true]{hyperref} 
	\hypersetup{
		colorlinks = true,
		linkcolor = blue,
		anchorcolor = blue,
		citecolor = blue,
		filecolor = blue,
		urlcolor = blue,
	}

\def\Loop{\ensuremath{\mathrm{Loop}}}
\def\Grph{\ensuremath{\mathbf{Grph}}}
\def\rGrph{\ensuremath{\mathbf{rGrph}}}
\def\Int{\ensuremath{\mathbf{Int}}}
\def\rgInt{\ensuremath{\mathbf{rgInt}}}
\def\rgIntSh{\ensuremath{\mathbf{rgIntSh}}}
\def\Set{\ensuremath{\mathbf{Set}}}
\renewcommand{\star}{*}

\def\rgIntSh{\ensuremath{\widetilde{\mathbf{rgInt}}}}
\def\IntSh{\ensuremath{\widetilde{\mathbf{Int}}}}
\def\Psh{\ensuremath{\mathrm{Psh}}}
\def\Hom{\ensuremath{\mathrm{Hom}}}
\def\Ob{\ensuremath{\mathrm{Ob}}}
\def\asSh{\ensuremath{\mathrm{asSh}}}

\def\RRp{\ensuremath{\mathbb{R}_{\geq 0}}}
\def\RR{\ensuremath{\mathbb{R}}}
\def\NN{\ensuremath{\mathbb{N}}}

\def\id{\ensuremath{\mathrm{id}}}

\def\Trp{\ensuremath{\mathrm{Tr}_p}}
\def\src{\ensuremath{\mathrm{src}}}
\def\tgt{\ensuremath{\mathrm{tgt}}}
\def\ids{\ensuremath{\mathrm{ids}}}
\def\Lt{\ensuremath{\mathrm{Lt}}}
\def\Rt{\ensuremath{\mathrm{Rt}}}

\def\op{\ensuremath{\mathrm{op}}}

\def\Yon{\ensuremath{\mathsf{Yon}}}

\def\LTS{\ensuremath{\mathsf{LTS}}}
\def\CDS{\ensuremath{\mathsf{CDS}}}
\def\CMP{\ensuremath{\mathsf{CMP}}}

\newcommand{\cat}[1]{\ensuremath{\mathbf{#1}}}

\def\A{\ensuremath{\mathbf{A}}}
\def\B{\ensuremath{\mathbf{B}}}
\def\C{\ensuremath{\mathbf{C}}}
\def\u{\ensuremath{\mathbf{u}}}
\def\x{\ensuremath{\mathbf{x}}}
\def\y{\ensuremath{\mathbf{y}}}

\def\dyn{\ensuremath{\mathrm{dyn}}}
\def\rdt{\ensuremath{\mathrm{rdt}}}

\newtheorem{thm}{Theorem}[section]
\newtheorem{defn}[thm]{Definition}
\newtheorem{prop}[thm]{Proposition}
\newtheorem{remark}[thm]{Remark}

\newcommand{\doublearrow}[2]{\begin{tikzcd}[ampersand replacement=\&]
	#1 \ar[r, shift left=+0.2em,"\src"] \ar[r, shift left=-0.2em,"\tgt"'] \& #2
	\end{tikzcd}}

\newcommand{\doublearrowids}[2]{\begin{tikzcd}[ampersand replacement=\&]
	#1 \ar[r, shift left=+0.6em,"\src"] \ar[r, shift left=-0.6em,"\tgt"'] \& \ar[l, "\ids" description] #2
	\end{tikzcd}}

\newacro{ACAS}{Aircraft Collision Avoidance System}
\newacro{RA}{Resolution Advisory}
\newacro{TA}{Traffic Advisory}
\newacro{ADS-B}{Automatic Dependent Surveillance - Broadcast}

\begin{document}

\title{Abstraction, Composition and Contracts:\\ A Sheaf Theoretic Approach\thanks{The authors of this paper were sponsored by NASA through the contract NNL14AA05C.}}

\author{Alberto Speranzon\thanks{Alberto Speranzon and  Srivatsan Varadarajan are with Honeywell Aerospace - Advanced Technology, Plymouth, MN. Emails: \texttt{firstname.lastname@honeywell.com}.} \and David I. Spivak\thanks{David I. Spivak is with the Department of Mathematics, MIT, Cambridge, MA. Email: \texttt{dspivak@math.mit.edu}.} \and Srivatsan Varadarajan$^\dagger$}

\maketitle

\begin{abstract}
	Complex systems of systems (SoS) are characterized by multiple interconnected subsystems. Typically, each subsystem is designed and analyzed using methodologies and formalisms that are specific to the particular subsystem model of computation considered --- Petri nets, continuous time ODEs, nondeterministic automata, to name a few. When interconnecting subsystems, a designer needs to choose, based on the specific subsystems models, a common abstraction framework to analyze the composition.
	
	In this paper we introduce a new framework for abstraction, composition and analysis of SoS that builds on results and methods developed in \emph{sheaf theory}, \emph{category theory} and \emph{topos theory}. In particular, we will be modeling behaviors of systems using sheaves, leverage category theoretic methods to define wiring diagrams and formalize composition and, by establishing a connection with topos theory, define a formal (intuitionistic/constructive) logic with a sound sheaf semantics.   
	
	To keep the paper more readable we will present the framework considering a simple but instructive example. 
	
%
%
\end{abstract}

\section{Introduction}

Most of today's complex systems are developed by first designing, analyzing and testing subsystems and then by interconnecting them. In order to achieve scalability without completely sacrificing analytical guarantees, one needs a formal framework to describe and analyze the composition that results from interconnecting subsystems. A key characteristic of such framework is the availability of methods to compose subsystems that adhere to different models of computation. Such models are conveniently used during the design and analysis process of each independent subsystem, however, heterogeneity typically leads to choose, on a case by case basis, a common abstraction model whose mathematical formalism strongly depends on such subsystems modeling choices. Furthermore, this decision is often -- as we see in the industrial world -- subjective and informal, leading to difficulties in re-usability.

Hybrid systems~\cite{Henzinger:96}, as a common abstraction model -- where discrete and continuous dynamics coexists -- have certainly demonstrated to be a successful paradigm for SoS modeling and analysis, see for example~\cite{PT-GP-PL:04, Frehse:05,RM-MZ:15} and references therein. In a setting where a finite state automaton and a dynamical system are used to model, for example, two different subcomponents, in order to compose the two systems within a hybrid system model, the designer will need to choose what the discrete and continuous state of the hybrid system are, define the correct transition rules (guards) and possible reset maps. This can be difficult for complex subsystems. In a different scenario, if subsystems were all modeled as automata, the designer would use a completely different common abstraction model.

Contract-based design methodologies~\cite{ASV-WD-RP:12,TH-SQ-SR-ST:02,PN-ASV-DB-LG-TV:15} tend to be more flexible in this context, since no explicit model of each subsystem is directly required to study the composition and only certain ``assume-guarantee'' contracts need to be defined. However, again, the mathematical formalization of each subcontract might be easily expressed using different formalisms, making the characterization of the contract of the composite system difficult to describe.

In what follows, we present an alternative approach for abstraction and composition of such systems and describe how such methods enable us to prove properties (or express contracts/requirements). The type of mathematical formalism suggested here, is related to early work by Ames~\cite{ames:06} and Tabuada~\cite{tabuada:08}. Here are taking a different approach, based on sheaf and topos theory. The payoff in using more abstract mathematics is a theory of abstraction and composition that supports a higher-order temporal logic, enabling us to express contracts, requirements, properties and verify them.

The paper is organized as follows: in Section~\ref{sec:sheaves} we introduce the key mathematical notions and in Section~\ref{sec:ACAS_example} we discuss their application to a problem in aerospace, namely a simplified \ac{ACAS}, inspired by the Traffic Collision Avoidance System currently used in aerospace. Section~\ref{sec:contracts} will discuss the link between the abstraction and a higher-order temporal logic. In Section~\ref{sec:conclusions} we draw the conclusions.

\section{Behaviors, Machines and Wiring Diagrams}
\label{sec:sheaves}
We will assume that the reader is familiar with category theory~\cite{Spivak:14,Awodey:10}. A deeper discussion on the theoretical foundations can be found in~\cite{DS-CV-PS:16} and~\cite{schultz-spivak:2017}.

\subsection{Behaviors As Sheaves}
\label{sec:interval_sheaves}
One of the central concepts that we use this paper is the notion of \emph{interval sheaves}. In order to define what these mathematical objects are, we first need to introduce the categories of \emph{continuous and discrete intervals}.
\begin{defn}
	The category of \emph{continuous intervals}, denoted with~$\Int$, is defined as follows:
		\begin{itemize}
			\item Objects $\Ob(\Int) = \{\ell \in \RRp\}$;
			\item Morphisms $\Hom_{\Int}(\ell',\ell) = \{\Trp | 0\leq p\leq \ell-\ell'\}$;
			\item Composition of morphisms: $\Trp \circ \mathrm{Tr}_p' = \mathrm{Tr}_{p+p'}$;
			\item Identity $\id_\ell = \mathrm{Tr}_0$ so that $\Trp \circ \id_\ell = \id_\ell \circ \Trp = \Trp$.
		\end{itemize}
\end{defn}
We can regard an object $\ell\in \Int$ to be the interval $[0,\ell] \subseteq \RRp$ and the morphism~$\Trp:\ell'\to\ell$ mapping an interval $[0,\ell']$ into the interval~$[0,\ell]$ by translation: $x\mapsto x+p$.

Replacing $\RRp$ by $\NN$, we can define the category~$\Int_N$ of \emph{discrete intervals}, where objects are natural numbers and morphisms are inclusions of smaller (discrete) intervals into larger ones.

Given the category $\Int$ we can consider an \emph{$\Int$-presheaf}, namely a functor $X:\Int^\op \to \Set$, where~$\Int^\op$ is the opposite category (where morphisms are the same as in $\Int$ but with the directions reversed, see~\cite{Spivak:14,Awodey:10}). For any continuous interval~$\ell \in \Int$ we refer to the elements $x \in X(\ell)$ as \emph{sections of~$X$ on~$\ell$}. Given a section $x \in X(\ell)$ and the map $\Trp: \ell' \to \ell$ we write $x|_{[p,p+\ell']}$ to denote the \emph{restriction map} $X(\Trp)(x) \in X(\ell')$. Similarly we define an $\Int_N$-presheaf as a functor $Y\colon\Int_N^\op\to\Set$.

\begin{figure}
	\centering
	\vspace*{-0.1cm}
	\ifdefined \doublecol
	\includegraphics[width=1.0\hsize]{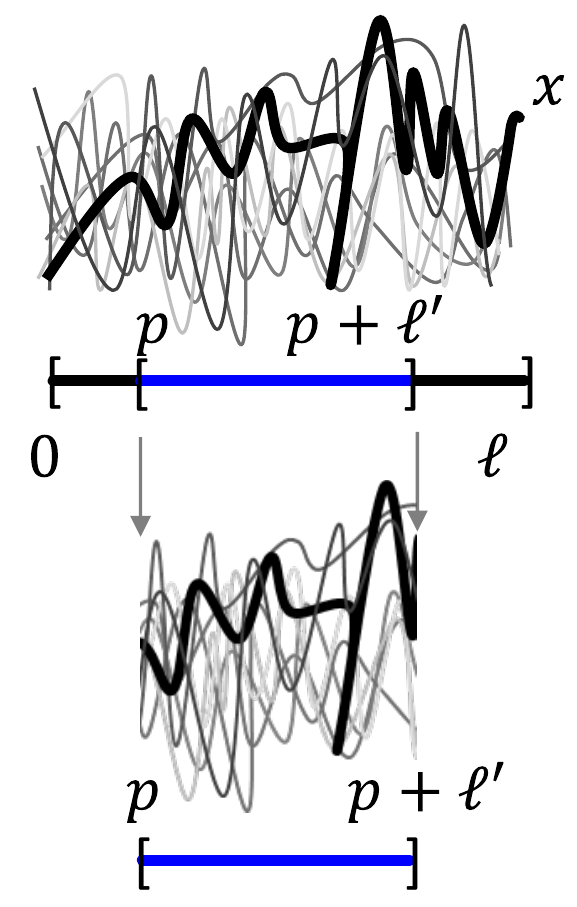}
	\else
	\includegraphics[width=0.25\hsize]{sheaf_example_2_cropped.pdf}
	\fi
	\caption{On top (thick solid line) a section $x \in X(\ell) \in \IntSh$ and on the bottom the its restriction $X(\Trp)(\ell) \in X(\ell')$.}\label{fig:example_presheaf}
\end{figure}

Intuitively, $\Int$-presheaves allows us to ``attach'' to an interval of length~$\ell$ ``arbitrary'' data structures\footnote{Although we have defined $\Int$-presheaves as a functor from~$\Int^{\op}$ to the category of sets,~$\Set$, one can generalize the definition to consider other categories with more structure, such as groups, rings, vector spaces, etc.}, modeling the systems' behaviors, as it will be clear later on. Figure~\ref{fig:example_presheaf} shows an example of a section of an $\Int$-presheaf and its restriction. Note that in this example, we are, in particular, ``attaching'' to  the interval the set of all continuous smooth signals and then picking one such `section'. A section of a similar $\IntSh_N$-presheaf would be represented as a discrete (sampled) signal.

%
%

Now, given two sections, $x_1 \in X(\ell_1)$, $x_2 \in X(\ell_2)$, we say that they are \emph{compatible} if the right endpoint of~$x_1$ matches the left endpoint of~$x_2$\footnote{The ``glueing'' in this paper will be represented as a compatibility at a single point, but this can be made more general, where one can consider compatibility over intervals and discontinuities~\cite{DS-CV-PS:16}.}. Intuitively, we should be able to ``glue'' two sections together if they are compatible. If this is the case, i.e.\ if for any compatible $x_1$ and $x_2$ there is a unique section $x_1 * x_2 \in X(\ell_1+\ell_2)$ whose right restriction  is~$x_2$ and whose left restriction is~$x_1$, then we say that the presheaf~$X$ satisfies the \emph{sheaf axiom}, see~\cite[Chapter 2]{Tenninson:75}, or equivalently that $X$ is a $\Int$-sheaf and write~$X\in \IntSh$. We can, in a similar fashion, define $\Int_N$-sheaves, the category which is denoted by~$\IntSh_N$. 

There is a functor $\asSh\colon\Psh(\Int) \to \IntSh$, called \emph{sheafification}, which freely adds a unique gluing for every pair of compatible sections. We will discuss sheafification in detail later using an example.

An important result that we will make use multiple times later, in the paper, is the following:
\begin{prop}{\cite[Proposition 3.2.2]{DS-CV-PS:16}}\label{prop:intN_graphs}
	There is an equivalence of categories, $\IntSh_N \simeq \Grph$, between the category of discrete interval sheaves and the category of graphs.
\end{prop}
The proof of this proposition can be found in~\cite{DS-CV-PS:16}. For clarity, we mention that an $n$-length section of an~$\Int_N$-sheaf are all the paths of length~$n$ over a graph, restrictions are sub-paths, and gluing is path concatenation.

In the following, we will extensively use the category of graph, $\Grph$, and the category of reflexive graphs, $\rGrph$. Although fairly standard, we have summarized in Appendix~\ref{apx:graphs} their definitions.

\subsection{Abstract Machines}

So far we have been generalizing the idea of behaviors over time intervals, with the property that if a behavior is well defined over an interval, it needs to be well defined for every subinterval. We now use this abstraction to model input/output systems, namely we define the concept of \emph{abstract machine}.

\begin{defn}{\cite[Section 4.1]{DS-CV-PS:16}}
	Let~$\tilde{I}, \tilde{O} \in \IntSh$ be interval sheaves. A \emph{$(\tilde{I},\tilde{O})$-machine} is the span:
	$$
		\begin{tikzcd}
			\tilde{I} & \tilde{S} \ar[r, "p^o"] \ar[l, "p^i"'] & \tilde{O}
		\end{tikzcd}
	$$
	where~$\tilde{S}$ is also an $\Int$-sheaf, and $p^i, p^o$ are sheaf morphisms, which we call the \emph{input} and \emph{output maps}. Equivalently, an abstract machine is the sheaf~$\tilde{S}$ together with the sheaf map $p\colon \tilde{S} \to \tilde{I} \times \tilde{O}$.
\end{defn}
It is interesting to point out that this representation is similar to the behavioral approach of systems provided by Willems~\cite{Willems:07-001}, which was recently considered, in the context of category theory, in~\cite{fong:16}.

The key important property of abstract machines is that they can be interconnected together to form new abstract machines.

\subsection{Wiring Diagrams}
\label{subsec:wiring_diagrams}
Interconnections of systems can be formalized as a category of \emph{wiring diagrams}. Formally, this is defined as a symmetric monoidal category $(\mathcal{W}_\mathcal{C},\oplus,0)$, where the objects are $\mathcal{C}$-labeled boxes (subsystems), representing the inputs, the outputs and their types. The type can be thought as the ``information'' carried by the input/output ports (or wires) and it is formalized as a sheaf. A morphism in the category $\mathcal{W}_\mathcal{C}$ is called the wiring diagram and it tells, as the words say, which outputs are being fed into which inputs. Details on these type of representation can be found in~\cite{vagner:15}. 

Given wiring diagram categories~$\mathcal{W}_\mathcal{C}$, we can consider a functor $F:\mathcal{W}_\mathcal{C} \to \mathbf{Cat}$, which we call $\mathcal{W}_\mathcal{C}$-algebra. 
If we think about interconnected systems, these will have a set of inputs/outputs feeding into each other (interconnection) and a few inputs/output that can be thought to connect to the external world. The functor~$F$, then enables to combine subsystems into a composite system in a way that respects the internal and external interconnections~\cite{vagner:15}.

Now, in the context of sheaves we have $\mathcal{C} = \Int_\bullet$\footnote{We use $\Int_\bullet$ to denote $\Int$ and $\Int_N$.}, namely the type of each input/output is a time-dependent signal.

\begin{prop}{\cite[Proposition 4.1.3 \& 4.4.3]{DS-CV-PS:16}} Abstract machines form a $\mathcal{W}_{\Int_\bullet}$-algebra.
\end{prop}

This means that if subsystems are abstract machines, and we have a wiring diagram representing the interconnections (among subsystems and the external world), we can compose them into a larger subsystem that retains the input/output compatibility with the external world.

\subsection{Explicit Construction of Abstract Machines}

In the following we define the construction of two important classes of abstract machines: continuous dynamical systems and labeled transition systems. 

\subsubsection{Continuous Dynamical System}

We define a continuous dynamical system as the tuple $F = (X, f^\dyn, f^\rdt, X_0)$, where~$X$ is a smooth manifold, $f^\dyn\colon I \times X \to TX$ are the dynamics, $I$ is the input space, $TX$ is the tangent bundle, $f^\rdt:S \to O$ is a smooth map, $O$ is the output space and $X_0\subseteq X$ is a set of initial states. In a more standard form we write:
\begin{align}
	\dot{x} &= f^\dyn(x,u)\,, \quad u\in I\,, x\in X\,, x_0\in X_0\,,\label{eq:state_trans_cds}\\
	y &= f^\rdt(x)\,, \quad y \in O\,.\label{eq:output_cds}
\end{align}
We can associate a sheaf~$\tilde{S} \in \IntSh$ to~$F$ by
\begin{align*}
\tilde{S}(\ell) =& \big\{(u,s) \colon [0,\ell] \to I \times X \mid u, x \text{ are smooth }\\&\text{ and } \dot{x} = f^\dyn(x,u)\big\}\,.
\end{align*}
Thus sections of the $\Int$-sheaf~$\tilde{S}$ are solutions to the differential equation~\eqref{eq:state_trans_cds} for a given set of inputs and an initial condition $x_0\in X_0$. Defining sheaves $\tilde{I}$ and $\tilde{O}$ with sections (trajectories) $[0,\ell]\to I$ and $[0,\ell]\to O$, the corresponding machine for the continuous dynamical system \eqref{eq:state_trans_cds}--\eqref{eq:output_cds} is the span
$$
\begin{tikzcd}[column sep=0.5cm]
&   {[0,\ell]} \ar[r] \ar[ldd,"\id"',pos=0.7]\ar[rdd,"\id",pos=0.7] & I \times X \ar[ldd,"\pi_1"',pos=0.7] \ar[rdd,"f^\rdt \circ \pi_2",pos=0.7] &  \\
&&&\\
{[0,\ell]}\ar[r] & I & {[0,\ell]}\ar[r] &  O
\end{tikzcd}
$$
where $\pi_1, \pi_2$ are projection functions such that $\pi_1\colon I \times X \to I$ and $\pi_2:I \times X \to X$, and where the top map is a section of the state of the machine.

It is possible, using similar arguments, to functorially assign an abstract machine to any discrete time dynamical system, whose dynamics are described by difference, instead of differential, equations. One just needs to pick an embedding: a fixed time step, or a random time-step, or a time-step dependent on some other parameter, etc.

\subsubsection{Labeled Transition System}

\begin{figure}[t]
	\centering
	\ifdefined \doublecol
	\includegraphics[width=0.7\hsize]{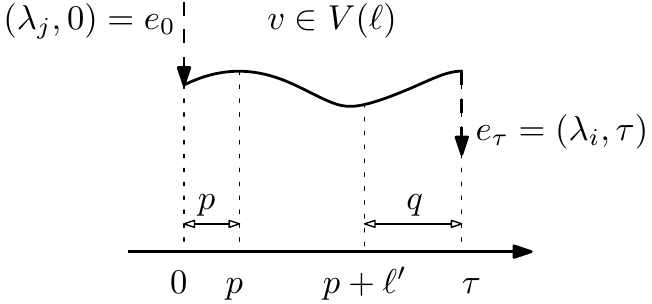}
	\else
	\includegraphics[width=0.45\hsize]{presheaf_section.pdf}
	\fi
	\caption{A $\ell$-length section of~$\bar{H}$.}\label{fig:presheaf_section}
\end{figure}

Another relevant construction that we will use in the paper is one that allows us to abstract a (labeled) transition system into a $\Int$-sheaf abstract machine. For this construction we need to define two functors,~$\Gamma$ and~$R$ such that
\begin{equation} 
\label{eq:chain_functors}
\begin{tikzcd}
\Grph \ar[r,"\Gamma"] & \rgIntSh \ar[r,"R"] & \IntSh
\end{tikzcd} 
\end{equation}
The objects in $\rgIntSh$%
\footnote{For a formal definition of~$\rgIntSh$, please see~\cite[Section A.2.1]{DS-CV-PS:16}. For the discussion here, one can think elements of~$\rgIntSh$ being either a ``transition'' or a ``vertex'', namely the basic components of a signal as the one shown in Figure~\ref{fig:presheaf_section}.}
are pairs $H = (V,G)$ where $V \in \IntSh$ and $G \in \rGrph$  and~$G=(V(0),E,\src,\tgt,\ids)$. We call~$H$ a \emph{hybrid sheaf datum}. The word ``hybrid'' here is, and is not, related the the classical concept of hybrid systems, see~\cite{RG-RS-AT:09} and references therein. It is related because, as for standard hybrid systems, there the need of modeling a mix of continuous and discrete behaviors, where the $\Int$-sheaf~$V$ captures the continuous behavior and the reflexive graph~$G$ models the discrete jumps. It is however, different than the classical hybrid automata models in~\cite{Henzinger:96,RG-RS-AT:09} as we are not prescribing the continuous behavior to be described by dynamical systems, like~$F$. The hybrid sheaf datum can be considered as a ``template'' that enables us to capture mixed --- continuous and discrete --- behaviors and it will enable us to represent specific dynamics only once it is used within an abstract machine framework. The example we will describe later in the paper will help to clarify this point further.

The functor~$\Gamma$ allows us to construct a hybrid sheaf datum from a graph, $\Gamma(G) = H = (V,G')$, where $G'\in \rGrph$ is the reflexive graph associated to~$G$, and the functor~$R$, which we call \emph{realization functor}, defines an~$\Int$-sheaf from it. We will use this to construct abstract machines, as we will show more explicitly when we discuss a specific example. 
The functor~$R$ is defined by first constructing the $\Int$-presheaf~$\bar{H}$ and then by sheafifying it. More specifically, we define the presheaf $\bar{H} \in \Psh(\Int)$ from~$H$ as follows: 
\begin{itemize}[leftmargin=*,topsep=0pt,itemsep=-1ex,partopsep=1ex,parsep=1ex]
	\item for 0-length sections we define $\bar{H}(0) \coloneqq E$;
	\item for $\ell$-length sections ($\ell >0$) we define the sections~$\bar{H}(\ell)$ by the following pullback in~$\Set$:
	\begin{equation}\label{eq:pullback_barH}
	\begin{tikzcd}
		\bar{H}(\ell) \ar[r] \ar[d]  \ar[dr,phantom, "\lrcorner", pos = 0.0, yshift=-0.5em] &  E \times E \ar[d,"\tgt \times \src"]\\
		V(\ell)  \ar[r,"{(\lambda_0,\rho_0)}"']  & V(0) \times V(0)
	\end{tikzcd}	
	\end{equation}
	where $\lambda_0, \rho_0:V(\ell)\to V(0)$ are the left and right endpoints (restrictions) of a $\ell$-length section. Sections $\bar{H}(\ell)$ can be thought as the behavior, in the interval~$[0,\ell]$ of the system at a certain vertex (or, with a better word, ``flow'') $v\in V(\ell)\in \IntSh$, together with two transitions (edges), one into~$v$ and the other out of~$v$. Figure~\ref{fig:presheaf_section}, shows an example of a section~$\bar{H}(\ell)$.

	To define restriction maps for~$\bar{H}$, let us consider~$\Trp\colon [0,\ell'] \mapsto [p,p+\ell']\subseteq[0,\ell]$ and define~$q \coloneqq \ell - (p+\ell')$, so that~$p$ is the length to the left of the subinterval and~$q$ is the length at the end, see Figure~\ref{fig:presheaf_section}. We then have two possible cases: $0 = \ell' < \ell$ and $0 < \ell' < \ell$. Note that the case $\ell' = \ell$ is trivial because the restriction in this case is just the identity. So we have that either $p \neq 0$ or $q \neq 0$ or both are non-zero. Assume~$\ell' = 0$ and let $(e_0,v,e_\ell) \in \bar{H}(\ell)$ be a section. We then define the restriction $\bar{H}(\Trp):\bar{H}(\ell) \to \bar{H}(0)$ as
	$$
	\bar{H}(\Trp)(e_0,v,e_\ell) = 
	\begin{cases}
		e_0 & \text{if $p = 0$, $q \neq 0$\,,}\\
		e_\ell & \text{if $p \neq 0$, $q = 0$\,,} \\
		\ids(v|_{[p,p]}) & \text{if $p \neq 0$ and $q \neq 0$}\,.
	\end{cases}
	$$
	Let~$\ell' >0$, let $(e_0,v,e_\ell) \in \bar{H}(\ell)$ be a section and let us define $e_0' = \ids(\lambda_0(v|_{[p,p+\ell']}))$, $e'_\ell = \ids(\rho_0(v|_{[p,p+\ell']}))$ the reflexive edges (loops) at the left and right endpoints. Then the restriction $\bar{H}(\Trp):\bar{H}(\ell) \to \bar{H}(\ell')$ is given by
	$$
	\bar{H}(\Trp)(e_0,v,e_\ell) = 
	\begin{cases}
	(e_0,v|_{[p,p+\ell']},e_\ell') & \text{if $p = 0$, $q \neq 0$\,,}\\
	(e_0',v|_{[p,p+\ell']},e_\ell) & \text{if $p \neq 0$, $q = 0$\,,} \\
	(e_0',v|_{[p,p+\ell']},e_\ell') & \text{if $p \neq 0$, $q \neq 0$}\,.
	\end{cases}
	$$
	Figure~\ref{fig:presheaf_restrictions} shows the three restrictions of an $\ell$-length to an $\ell'$-length section.
	\begin{figure}[t]
		\centering
		\ifdefined \doublecol
			\includegraphics[width=0.8\hsize]{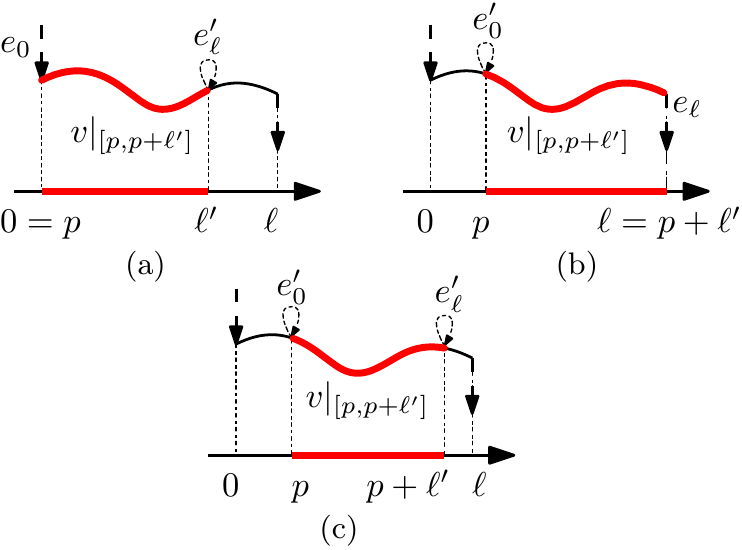}
		\else
			\includegraphics[width=0.65\hsize]{presheaf_restriction.pdf}
		\fi
		\caption{Restrictions of a $\ell$-length section to a $\ell'$-section for various values of~$p$ and~$q$.}\label{fig:presheaf_restrictions}
	\end{figure}
\end{itemize}
Given the $\Int$-presheaf~$\bar{H}$, we need to sheafify it to obtain a $\Int$-sheaf. We thus define $R(H)\coloneqq\asSh(\bar{H})$. 

\subsection{Composition}
\label{subsec:composition}
As mentioned in~\ref{subsec:wiring_diagrams}, there is a formula for interconnecting dynamical systems (or more generally machines) according to any wiring diagram. This formula is fully associative, meaning we can chunk the diagram in arbitrary ways, allowing us to zoom in and out. The fact that it operates on machines, i.e.\ spans of $\Int$-sheaves, means that the composition is very general; in particular, we can compose discrete systems and continuous systems by considering both in terms of $\Int$-sheaves.
%
%

While the general composition formula is beyond the scope of the present paper (see \cite{DS-CV-PS:16} for full details), it is essentially a matter of pullbacks in the category of $\Int$-sheaves. Machines are composed by sharing variables wherever they are interconnected; a state for the composite machine consists of a state for each component, such that the inputs and outputs agree on shared wires. For example, given two machines: $\mathcal{M}_1\colon \tilde{I}_1 \leftarrow \tilde{S}_1 \to \tilde{O}_1$ and $\mathcal{M}_2\colon \tilde{I}_2 \leftarrow \tilde{S}_2 \to \tilde{O}_2$, and a sheaf morphism $g\colon\tilde{O}_1\to\tilde{I}_2$, their composite is defined by the pullback shown below
$$
\begin{tikzcd}[column sep=0.01cm,row sep=0.05cm]
&&& \tilde{S}_\CMP \ar[rrd,"\pi_2"] \ar[lld,"\pi_1"'] \ar[dd,phantom,"\lrcorner", pos = 0.1, rotate = 90] &&& &&& & \tilde{S}_\CMP \ar[rddd,"p^o_2 \circ \pi_2"] \ar[lddd,"p^i_1 \circ \pi_1"']  &\\
& \tilde{S}_1 \ar[ldd,"p^i_1"'] \ar[rdd, "p^o_1"]&&&& \tilde{S}_2 \ar[ldd,"p_2^i"'] \ar[rdd, "p^o_2"]& && \simeq&& &\\
&&& \tilde{O}_1 \ar[ld,equal] \ar[rd, "g"] &&& &&&& &\\
\tilde{I}_1 && \tilde{O}_1 && \tilde{I}_2 && \tilde{O}_2 &&& \tilde{I}_1 & &  \tilde{O}_2
\end{tikzcd}
$$

\section{Example: Aircraft Collision Avoidance System (ACAS)}
\label{sec:ACAS_example}

To ground the discussion presented in the previous section we will show how we can apply the idea of $\Int$-based abstract machines to an \ac{ACAS} system. We take inspiration from the Traffic Collision Avoidance System (TCAS) II~\cite{TCAS:11}, which is present on airplanes to avoid in-air collisions. We have included a slightly more detailed description of an \ac{ACAS} system in the Appendix~\ref{apx:acas} for a reader who might be not familiar with it. 

In particular, the problem we consider is oversimplified -- for example it neglects the fact that TCAS II provides different climbing/descending rates -- as the intent is to show how we can leverage the idea of abstract machines and composition through wiring diagrams to formally construct a system from a set of subsystems. The lack of tools at this stage prevents us from tackling more complicated scenarios, however, as it will be clear from the modeling, abstraction and composition, the process can ported into an algorithm.

As described in Appendix~\ref{apx:acas}, the \ac{ACAS} can be decomposed into three subsystems: a collision logic (that decides which aircraft should climb and which should descend), a pilot that executes the maneuver and an aircraft that, based on the pilot's action, will change altitude. More specifically: (1) we model the \ac{ACAS} as a labeled transition system that is receiving altitude and maneuver information periodically, with period~$\tau$ from the other vehicle and outputs the maneuver the pilot should take to avoid collision; (2) we model the aircraft longitudinal dynamics as a continuous time system whose input is the elevator deflection angle; (3) we model the human as a map from maneuver to elevator angle. We will assume that the human delay is negligible at the time scales we are considering, although delays can be incorporated\cite{DS-CV-PS:16}.

Although the over model is clearly very simple, it will demonstrate that we can compose two subsystems, one modeled as a synchronous periodic discrete system and the other as an asynchronous continuous time dynamical system, by using sheaf-based abstract machines. 

It is well known that such a system can be modeled using hybrid systems, but as we mentioned in the introduction, this would require the designer to make a decision, upfront on what the continuous and discrete states are. Undeniably simple in this case, but for a larger system, this might not be the case. The point we make here is that we can formalize the abstraction and composition without requiring the designer to make such decisions upfront. 

\begin{figure}[h]
	\centering
	\ifdefined \doublecol
	\includegraphics[width=0.95\hsize]{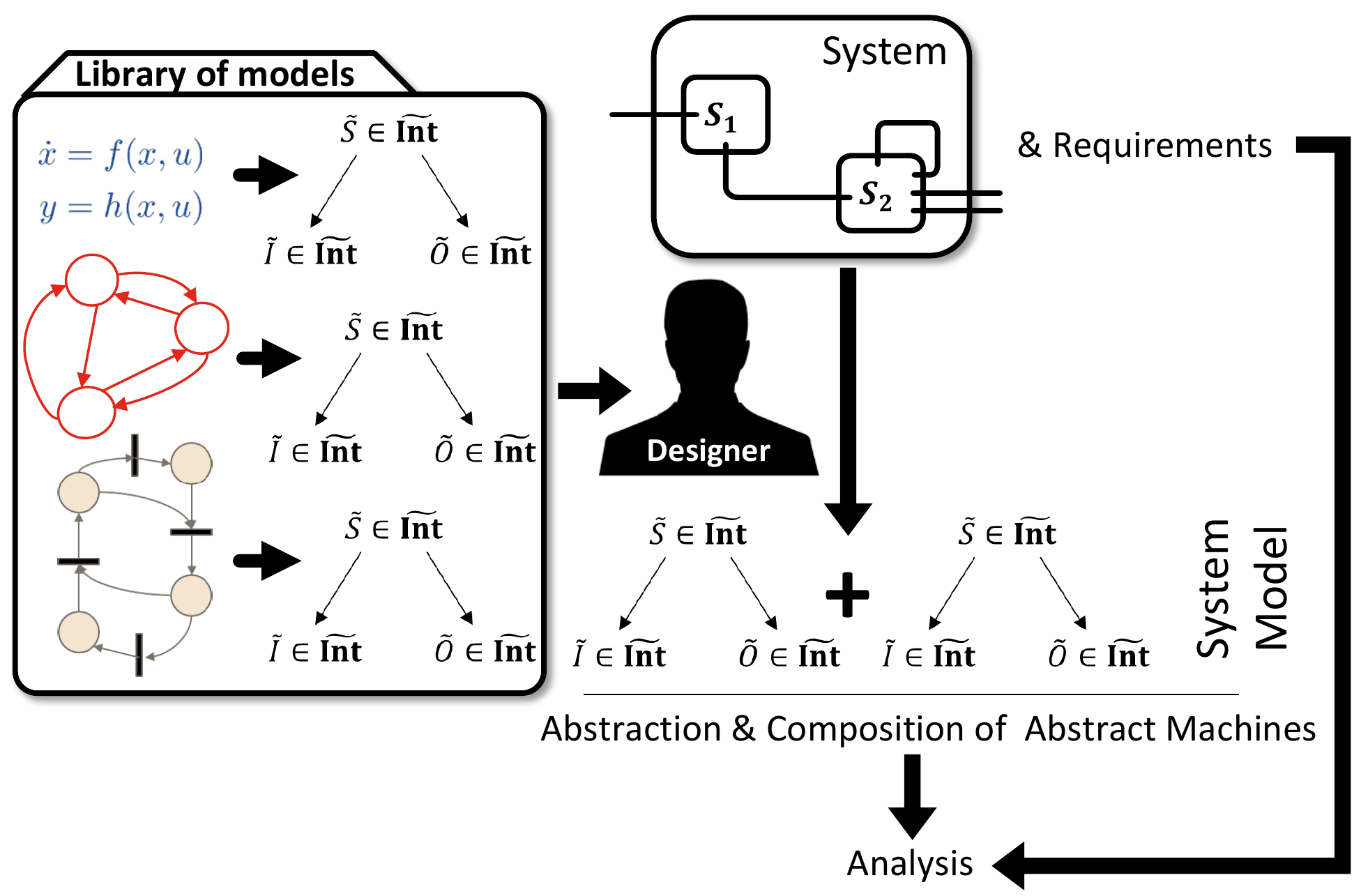}
	\else
	\includegraphics[width=0.7\hsize]{idea_cropped.pdf}
	\fi
	\caption{Overall idea of the proposed paradigm: from various models to a ``universal model'' of abstract machines enabling formal methods for abstraction, composition and analysis.}\label{fig:overall_idea}
\end{figure}

To ``see'' the potential advantage of the framework, without being distracted by the simplicity of the problem, we refer the reader to Figure~\ref{fig:overall_idea}. The idea is that models can be abstracted into the common language of sheaves and abstract machines, and through the wiring diagram algebra we can formally compose very different system models. We believe the effort in abstracting models into abstract machines is offset by a formal methodology fors modeling, composition and analysis.

In the next section, we will mostly discuss though the \ac{ACAS} example the path from the interconnected subsystems to the abstraction and composition. We will discuss after that the path from requirements to analysis. 

\subsection{ACAS Logic}

The \ac{ACAS} logic selects, based on the maneuver and altitude of the two aircrafts, three maneuvers: $M^i= \{\text{Climb},\allowbreak \text{Descend},\allowbreak \text{Level Flight}\}$, with $i \in \{1,2\}$. Let $A^i \in \RR$  be the altitude of the two vehicles and~$\delta \in \RRp$ be the minimal altitude difference between two aircraft so that no \ac{RA} is issued.   

We model the \ac{ACAS} as a labeled transition $\mathcal{T} = (S, \Lambda, \Omega, T, O, s_0)$ where:
\begin{itemize}[topsep=0pt,itemsep=-1ex,partopsep=1ex,parsep=1ex]
	\item $S = \{s_0, s_1 \dots, s_n\}$ is a finite set of state labels;
	\item $\Lambda =\{\lambda_1, \dots,\lambda_m\}$ is a finite set of input labels;
	\item $\Omega = \{\omega_1,\dots, \omega_n\}$ is a fine set of output labels;
	\item $T:\Lambda \times S \to S$ is a state transition map;
	\item $O:S \to \Omega$ is an output map;
	\item $s_0 \in S$ is the initial state.
\end{itemize}

\begin{center}
	\small
	\begin{tikzpicture}[>=stealth',shorten >=1pt,auto,node distance=3cm,every node/.style={inner sep=3,outer sep=3}]
	\node[initial,state,text width=25pt, align=center] (S)     {Level flight};
	\node[state,minimum size=40pt]         (q1) [below right of=S] {Climb};
	\node[state,minimum size=40pt]         (q2) [below left of=S]  {Descend};
	\path[->] (S)  edge [loop above, min distance=8mm] node {$\lambda_1:|A^1-A^2|\geq \delta$} (S);
	\path[->]	(S)	 edge [bend left]  node[align=center, near start, xshift=-0.5cm] {$\lambda_3:$\\$|A^1-A^2|< \delta\: \wedge$\\ $M^1= \text{`Descend'}$} (q1);
	\path[->]	(S)	 edge [bend right, left] node[near end, align=center] {$\lambda_2:$\\$|A^1-A^2| < \delta\: \wedge$\\ $M^1 = \text{`Climb'}$} (q2);
	\path[->]	(q1) edge  node[pos=0.7,xshift=0.25cm, align=center] {$\lambda_4,\lambda_5:$\\$|A^1-A^2|>\delta$} (S);
	\path[->] (q2) edge [right] node[pos=0.0, align=center] {} (S);
	\path[->] (q1) edge [loop right, min distance=5mm] node {$\lambda_3$} (q1);
	\path[->] (q2) edge [loop left, min distance=5mm] node {$\lambda_2$} (q2);
	\end{tikzpicture}
\end{center}

The set of states is $S = \{\text{'Level Flight'}, \allowbreak \text{'Climb'}, \allowbreak \text{'Descend'}\}=\{s_1,s_2,s_3\}$ and the set of output labels is such that $\Omega \equiv S$ and the initial state is $s_0 = s_1 = \text{'Level flight'}$.

The input label set $\Lambda = \{\lambda_1,\lambda_2,\lambda_3,\lambda_4,\lambda_5\}$ are the labels on the edges of the following diagram. To be more precise the inputs are tuples $(M^1,A^1,M^2,A^2)(t)$ provided at time~$t=k\tau$, $k\in\NN$, which determine the value of~$\lambda_i \in \{\mathrm{True},\mathrm{False}\}$.

The transition map~$T \subseteq S \times \Lambda$ is clear from the previous diagram.

For the model of interest we assume that the two aircrafts communicate to each other in a synchronous fashion using the \ac{ADS-B} (i.e. we assume the clocks to be all synchronized on GPS-time) and that the $i$-th aircraft has instantaneous access to its internal state $(M^i,A^i)$, and will receive instantaneously from the other aircraft its state. 

To model the \ac{ACAS} as an abstract machine, as defined in Section~\ref{sec:interval_sheaves}, we proceed in two steps: (1) we first model the signals associated to input, output and state as graphs~$\Int_N$-sheaves (see Proposition~\ref{prop:intN_graphs}) and (2) we apply the functors~\eqref{eq:chain_functors} to map $\Int_N$-sheaves to $\Int$-sheaves, as we are interested to build a continuous (common) abstraction.

The input signal is just a $\tau$-periodic sequence of events $\lambda_{k_1}, \lambda_{k_2}, \dots$. Such a discrete sequence of input events (where we abstract time away) can be modeled as an $\Int_N$-sheaf, and its equivalent representation, as a graph, is a loop graph $\Loop(\Lambda) = (\Lambda \rightrightarrows v_\star) \in \cat{Grph}$, where we associate to each self-loop, labeled by~$\lambda_i$, a single vertex~$v_\star$ (as the input signal is not defined in between two instances at which events occur).

The output of the transition system~$\mathcal{T}$ is a piecewise constant signal of period~$\tau$ modeling the fact that the \ac{ACAS} decision is persistently provided to the pilot and updated every~$\tau$ seconds. We use an $\Int_N$-sheaf to model the sequence of outputs, where we abstract the time away and thus treat the output as a sequence of labels. In this case we can use a complete graph $\mathcal{K}(\Omega) = (\Omega \times \Omega \rightrightarrows \Omega) \in \cat{Grph}$ as model. More specifically, we have that any output sequence can be seen as a path over a complete graph whose vertices are the output symbols. Note that~$\mathcal{K}(\Omega)$, as defined above, will in general produce all possible output sequences formed by an arbitrary concatenation of~$\omega_i$'s. This includes also sequences that are \emph{not valid} with respect to the transition system~$\mathcal{T}$, e.g. any output of the type~$\cdots\,\omega_2\omega_3\,\cdots$. The reason for this, is that we still need to couple the inputs to the outputs through the state. 

Finally, to model the state evolution of~$\mathcal{T}$ we consider, again, a graph $\mathcal{G}(\Lambda,S)= (\Lambda\times S \rightrightarrows S) \in\cat{Grph}$, where $\tgt(\Lambda,S) = T(\Lambda,S)$ and $\src(\Lambda,S) = \pi_2$. 

\begin{figure}[t]
	\centering
	\ifdefined \doublecol
		\includegraphics[width=0.8\hsize]{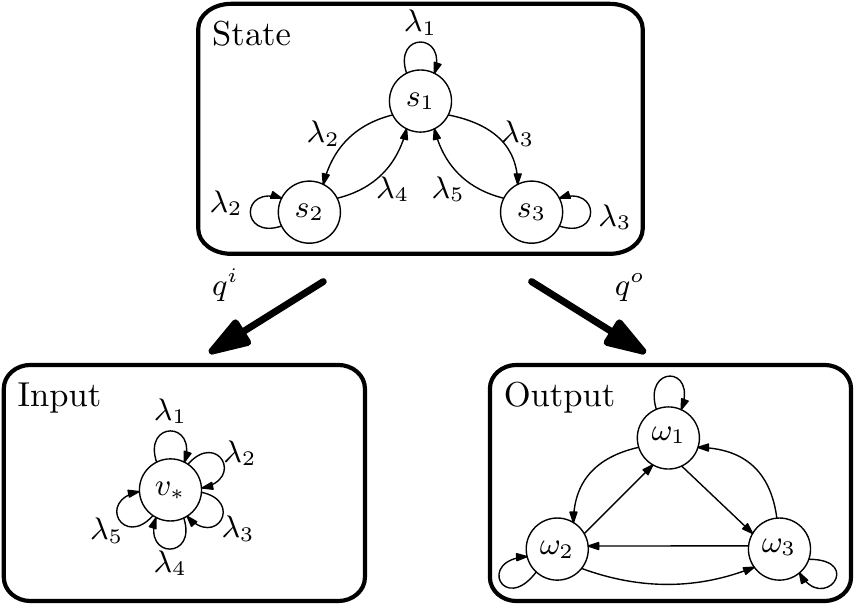}
	\else
		\includegraphics[width=0.6\hsize]{LTS_intN.pdf}
	\fi
	\caption{$\IntSh_N$ abstract state machine representing the transition system~$\mathcal{T}$. The maps~$q^i$ and~$q^j$ are defined in~\eqref{eq:input_output_state_model}.}\label{fig:abstract_machine}
\end{figure}

Putting these models together we obtain the abstract model in Figure~\ref{fig:abstract_machine} that is an~$\Int_N$-sheaf representation of the label transition system.
Mathematically we can represent the $\Int_N$-based abstraction using the following commutative diagram:
\begin{equation}\label{eq:input_output_state_model}
\begin{tikzcd}[column sep=0.02cm,row sep=0.35cm]
	& \Lambda \times S \ar[ldd, "\pi_1"',pos=0.3] \ar[rrrr, shift left=-0.2em,"T"'] \ar[rrrr,shift left=0.2em,"\pi_2"] \ar[rrrrdd,"h",pos=0.80] &&&& S \ar[lllldd,"g"',pos=0.80]\ar[rrdd,"O", pos=0.3] &&&&& \text {State} \ar[ldd, "{q^i =(\pi_1,g)}"'] \ar[rdd,"{q^o =(h,O)}"] &&\\
	&&&&&&& \iff &&&&\\
	\Lambda \ar[r, shift left=-0.2em] \ar[r,shift left=0.2em]  & v_\star &&&&\Omega \times \Omega \ar[rr, shift left=-0.2em,"\pi_2"'] \ar[rr,shift left=0.2em,"\pi_1"] && \Omega &&\text{Input}&& \text{Output}
\end{tikzcd}
\end{equation}
where $h:\Lambda \times S \rightarrow \Omega \times \Omega: (\lambda,s) \mapsto (O(s), (O(T(\lambda,s))))$ and~$g:S \to \{v_\star\}$ is the trivial map that maps every element of~$S$ to~$v_\star$.

Given the machine~\eqref{eq:input_output_state_model}, defined in~$\IntSh_N$, we are now going to apply the functors~\eqref{eq:chain_functors} to transform the machine in Figure~\ref{fig:abstract_machine} into an \emph{abstract machine}. We proceed by first transforming the input, output and state and then finally the maps~$q^i$ and$q^o$.

\subsubsection{Modeling of the Input as~$\IntSh$} 

Given~$\Loop(\Lambda) \in \Grph$ the first step is to apply the functor~$\Gamma$ in~\eqref{eq:chain_functors} to obtain a hybrid sheaf datum, which in some sense introduces time into the model. Let $H(I)\coloneqq\Gamma(\Lambda)$ so $H_I = (V_I,G_I)$ where $V_I = \{v_\star\} \times \Yon_\tau \simeq \Yon_\tau$,%
\footnote{$\Yon_\tau\in\IntSh$ is the Yoneda, or representable sheaf, for $\tau\in\Int$. It is defined by $\Yon_\tau(\ell)\coloneqq \{p \in \RRp | p \leq \tau - \ell\}$. Note that $\Yon_\tau(\ell) = \emptyset$ for~$\ell > \tau$.}
so that $V(0) = \{v_\star\} \times \Yon_\tau(0) \simeq \Yon_\tau(0)$. Intuitively,  sections of~$V_I(\ell)$ are length-$\ell$ subintervals of $[0,\tau]$. Now,~$G_I$ is a reflexive graph constructed from~$\Loop(\Lambda)$ by taking~$V_I(0)$ as vertices and $E_I = \Lambda \sqcup V_I(0)$ as edges. One can think of an edge as either a input label~$\lambda_i$ or a time-instant (vertex).

\begin{figure}
	\centering
	\vspace*{-0.3cm}
	\ifdefined \doublecol
	\includegraphics[width=1.0\hsize]{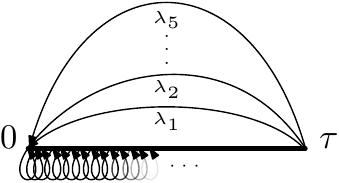}
	\else
	\includegraphics[width=0.4\hsize]{input_reflx_graph.pdf}
	\fi
	\caption{Visualization of the reflexive graph~$G_I$. There are infinite many self-loops at every time/vertex~$v$.}\label{fig:rflx_graph_input}
\end{figure}

A visualization of~$G_I$, to clarify better the object we are constructing, is shown in Figure~\ref{fig:rflx_graph_input}. The vertex set is clearly infinite (every point in the continuous interval is a vertex), and we have a self-loop for every vertex (time-instant) and edges labeled by~$\lambda_i$ from the vertex~$\tau$ to~$0$.

Mathematically we have~$G_I$ defined by the diagram:
$$
\begin{tikzcd}[column sep=0.05cm]
	G_I:  E_I =& \Lambda \ar[rrrr, "\src'(\lambda_i)=\tau" description,  bend left=45] \ar[rrrr, "\tgt'(\lambda_j) = 0" description, bend left=25] & \bigsqcup  & V_I(0) \ar[rr,"\src'(v) = v"'  description, bend right=40],  \ar[rr,"\tgt'(v)=v" description, bend right=20] &\hspace*{2cm}& V_I(0)  \ar[ll,"\ids(v) = v"'  description]
\end{tikzcd}
$$
Given an $\rgInt$-sheaf we can derive the~$\Int$-presheaf~$\bar{H}_I(\ell) \in \Psh(\Int)$, as the pullback~\eqref{eq:pullback_barH} where $E = E_I$. Thus we have that sections of $\bar{H}_I(\ell)$ are signals as the the one depicted in Figure~\ref{fig:presheaf_section}, but that are constant over the interval. The discrete transitions occur at the endpoints of the interval.

The restriction maps of the presheaf~$\bar{H}_I$, defined in~Section~\ref{sec:interval_sheaves}, are specialized for the input below. Let $x = (e_0, v,\allowbreak e_\tau)\allowbreak \in\allowbreak E_I \times V_I(\ell) \times E_I$ be a section, then we have restrictions 
\begin{align*}
	&\bar{H}_I(\Trp)((\lambda_j,0),v,(\lambda_i,\tau)) =\\
	& \begin{cases}
		(\ids(v_{\Lt}), v,(\lambda_i,\tau)) & \text{if $p \neq 0$ and $q = 0$,}\\
		((\lambda_j,0), v, \ids(v_{\Rt})) & \text{if $p =0$ and $q \neq 0$,}\\
		(\ids(v_{\Lt}),v,\ \ids(v_{\Rt})) & \text{if $p\neq0$ and $q\neq0$,}
	\end{cases}
\end{align*}
where $v\in \Yon_\tau(\ell)$, where~$v_\Lt = v|_{[p,p]}$ and~$v_\Rt=v|_{[p+\ell',p+\ell']}$ and where $\ids(v_{\Lt}) = v_{\Lt}$ and $\ids(v_\Rt) = v_\Rt$. These restrictions are as the one shown in~Figure~\ref{fig:presheaf_restrictions} with, again, the difference being that the red sections are constant (all taking value~$v_\star$). 

Given the $\cat{Int}$-presheaf~$\bar{H}_I$, we are going to use the realization functor~$R$ in~\eqref{eq:chain_functors} to obtain the sought model in~$\IntSh$. As we mentioned in~Section~\ref{sec:interval_sheaves}, $R \equiv \asSh$, namely the sheafification functor. We describe next how this acts on~$\bar{H}_I$.  

Consider the following set $\mathcal{L}_\ell(n) = \{(\ell_1,\ell_2,\dots,\ell_n) | \ell_i \geq 0,\sum_i \ell_i = \ell\}$. Then, $\tilde{I} = R(H_I) \in \IntSh$ has sections
\begin{align*}
	&\tilde{I}(\ell) =\{(\ell_1,\ell_2,\dots,\ell_n, x_1, x_2,\dots,x_n)| \\ &(\ell_1,\ell_2,\dots,\ell_n) \in \mathcal{L}_\ell(n),
	x_i\in \bar{H}_I(\ell_i), x_i|_{\Rt} = x_{i+1}|_{\Lt}\}\big / \sim\,,
\end{align*}
where we say that $(\ell_1,\dots,\ell_n) \sim (\ell'_1,\dots,\ell'_m)$ with $(\ell_1,\dots,\ell_n) \in \mathcal{L}_\ell(n)$ and $(\ell'_1,\dots,\ell'_m) \in \mathcal{L}_\ell(m)$ if and only if $(\ell'_1,\dots,\ell'_m)$ is a refinement of $(\ell_1,\dots,\ell_n)$.
\begin{figure}[t]
	\centering
	\ifdefined \doublecol
		\includegraphics[width=1\hsize]{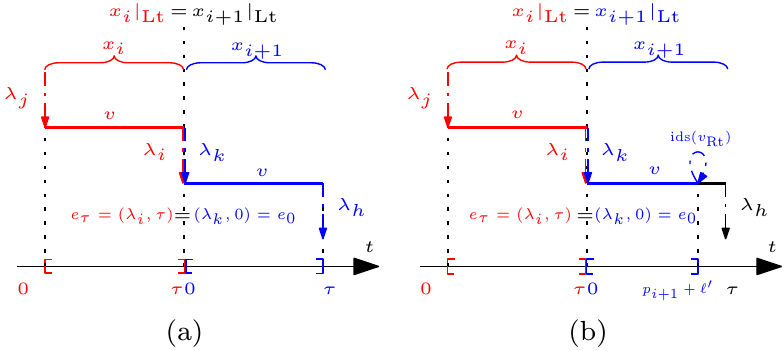}
	\else
		\includegraphics[width=0.7\hsize]{input_sheaf.pdf}
	\fi
	\caption{``Gluing'' of sections of~$\bar{H}_I$.}\label{fig:sheafification_1}
\end{figure}		
We have that the following main cases:
\begin{enumerate}[leftmargin=*,topsep=1pt,itemsep=-1ex,partopsep=1ex,parsep=1.5ex]
	\item $q_i = 0$ and $p_{i+1} = 0$ with $p_i$ and~$q_{i+1}$ arbitrary. In this case we have that sections glue as shown in Figure~\ref{fig:sheafification_1}(a)-(b), where in the first case we have $q_{i+1} = 0$ and in the second $q_{i+1} \neq 0$. 

	The other cases are very similar where the self-loop appears on the~$x_i$ or both. 
	
	\item $q_i \neq 0$ and $p_{i+1} \neq 0$ with $p_i$ and~$q_{i+1}$ arbitrary. In this case the gluing will happen by ``aligning'' self-loops together as shown in Figure~\ref{fig:sheafification_2}(a)-(b) for the case where $p_i =0$, $q_{i+1} = 0$, in (a), and $q_{i+1} \neq 0$, in (b).
	\begin{figure}[tb]
		\centering
		\ifdefined \doublecol
			\includegraphics[width=0.80\hsize]{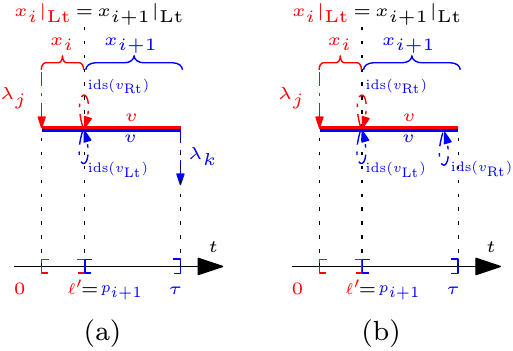}
		\else
			\includegraphics[width=0.46\hsize]{input_sheaf_2.pdf}
		\fi
		\caption{``Gluing'' of sections of~$H_I$.}\label{fig:sheafification_2}
	\end{figure}	
	It is not difficult to see that in this case the sheafification will produce one of the four possible sections: Figure~\ref{fig:presheaf_section} and Figure~\ref{fig:presheaf_restrictions}.
\end{enumerate}
The~$\Int$-sheaf $\tilde{I}$ we have build through this process is our model for input. We will next briefly discuss the output (while the state is discussed in Appendix~\ref{apx:state}). We will use the same process as above and thus discuss only the main differences.

\subsubsection{Modeling the output as~$\IntSh$}

Associated to the complete graph~$\mathcal{K}(\Omega)$ we construct the  $\Int$-sheaf $V_O = \Omega \times \Yon_\tau$ and the set $E_O = (\Omega \times \Omega) \sqcup V_O(0)$, where $V_O(0) = \Omega \times \Yon_\tau(0)$, to build the hybrid sheaf datum for the output,~$H_O=(V_O,G'_O)$. We have that $G'_O$ is the following reflexive graph:
$$
\begin{tikzcd}[column sep=0.05cm]
	G'_O:\hspace*{-0.2cm} &\Omega \times \Omega\hspace*{-0.25cm} 
	\ar[rrrr, "{\src'(\omega_i,\omega_j)=(\src(\omega_i,\omega_j),\tau)=(\omega_i,\tau)}" description,  bend left=45] \ar[rrrr, "{\tgt'(\omega_i,\omega_j)=(\tgt(\omega_i,\omega_j),0)=(\omega_j,0)}" description, bend left=20] & \bigsqcup\hspace*{-0.25cm}  & \Omega \times \Yon_\tau(0) 
	\ar[rr,"{\src'(\omega,p)=(\omega,p)}"'  description, bend right=45],  
	\ar[rr,"{\tgt'(\omega,p)=(\omega,p)}" description, bend right=20] &\hspace*{2cm}& \Omega \times \Yon_\tau(0)  \ar[ll,"{\ids(\omega,p)=(\omega,p)}"' description]
\end{tikzcd}
$$
The the presheaf $\bar{H}_O(\ell) = V_O(\ell) \times_{V_O(0) \times V_O(0)} E_O \times E_O$, obtained by the pullback~\eqref{eq:pullback_barH} has sections representing piecewise constant signals whose constant value depends on the output label~$\omega_i$.
	
The restriction maps are, for $s \in [p,\tau-q]$:
\begin{align*}
&\bar{H}_O(\Trp)(((\omega_i,\omega_j),0),(\omega_j,s),((\omega_i,\omega_k),\tau)) = \\
&\begin{cases}
(\ids((\omega_j,p)), (\omega_j,s),((\omega_i,\omega_k),\tau) & \text{if $p \neq 0$ and $q = 0$,}\\
((\omega_i,\omega_j),0), (\omega_j,s), \ids((\omega_j,q))) & \text{if $p =0$ and $q \neq 0$,}\\
(\ids((\omega_j,p)),(\omega_j,s), \ids((\omega_j,q))) & \text{if $p\neq0$ and $q\neq0$.}
\end{cases}
\end{align*}

To obtain a $\Int$-sheaf we need to sheafify the presheaf~$\bar{H}_O$ by applying the realization functor~$R$. The sections of the output $\Int$-sheaf, $\tilde{O}$, following the same argument as for the input, will be signals as the one shown on the bottom right of Figure~\ref{fig:input-output-sheaf-maps}. 

As the mechanics to obtain the $\Int$-sheaf representation for the state is a very similar exercise, we refer the reader to the Appendix~\ref{apx:state}.

\subsubsection{Abstract State Machine Representing the ACAS}

Now, given the input, output and state $\Int$-sheaves, we can build the following abstract machine
$$
\begin{tikzcd}
	\tilde{I}_\LTS \in \IntSh & \tilde{S}_\LTS \in \IntSh  \ar[l, "p^i"'] \ar[r,"p^o"]&\tilde{O}_\LTS\in \IntSh
\end{tikzcd}
$$
where use the subscript $\LTS$ to specifically indicate that we are considering the Labeled Transition System. The maps~$p^i$ and~$p^o$ are the input and output sheaf maps.

The input sheaf~$p^i:\tilde{S}_\LTS \to \tilde{I}_\LTS: \mathcal{S}(\ell) \mapsto \mathcal{I}(\ell)$ for any $\ell \in \RRp$, where~$\mathcal{S}(\ell)$ and~$\mathcal{I}(\ell)$ are state and input sections respectively, is defined as~$R(q^i)$, where~$R$ is the realization functor.

Let $x_s=(\ell_1,\dots,\ell_n,x_{s_1},\dots,x_{s_n}) \in \tilde{S}_\LTS(\ell)$ and $x_i=(t_1,\dots,t_m,x_{i_1},\dots,x_{i_m})\in \tilde{I}_\LTS(\ell)$ be $\ell$-length sections of the state and input. First note that if we have two partitions of the $\ell$-interval, $(\ell_1,\dots,\ell_n)$ and $(t_1,\dots,t_m)$, we can consider the coarser one of the two to define the state-to-input map~$p^i$, given that sections of the state and input are defined up to refinements. Without loss of generality let us assume this is $(\ell_1,\dots,\ell_n)$. Now within a subinterval~$\ell_k$ the state~$x_{s_k}$ is represented as a sequence of ordered transitions and vertices and thus we need to define how the map~$p^i$ behaves for these and how it maps to $x_{i_k}$. 

\begin{figure}[H]
	\centering
	\ifdefined \doublecol
	\includegraphics[width=0.95\hsize]{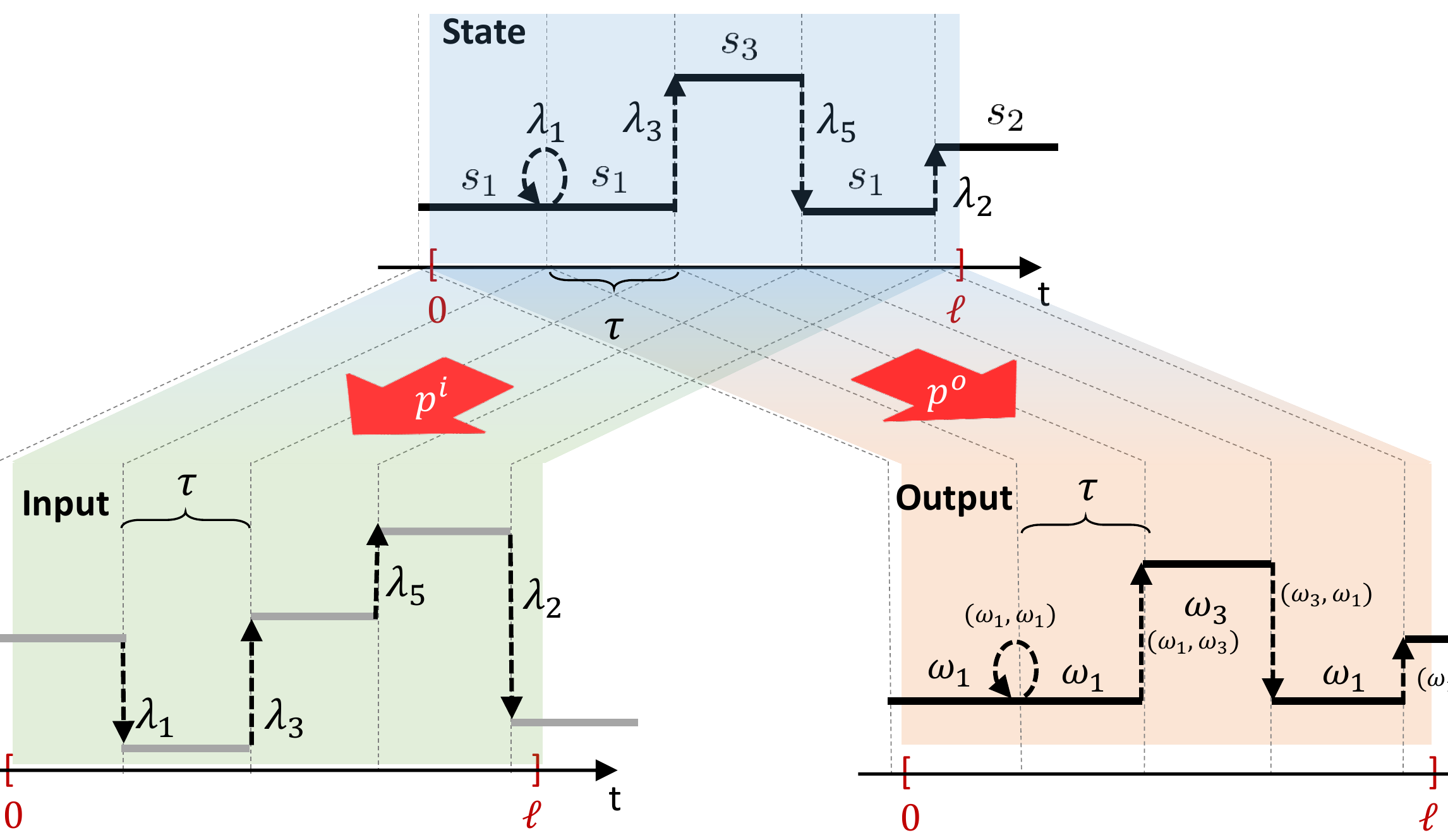}
	\else
	\includegraphics[width=0.8\hsize]{input_ouptut_state_model.pdf}
	\fi
	\caption{On the top a $\ell$-length section of the state sheaf and on the left the associated input sheaf through the state-to-input sheaf map~$p^i$. On the right the associated output sheaf through the state-to-output sheaf map~$p^o$. Note that the input and output sheaf maps ensure that the signals are ``aligned''.}\label{fig:input-output-sheaf-maps}
\end{figure}

Note first that, given the composition properties of~$q^i$ and~$q^o$, in~\eqref{eq:input_output_state_model}, we have that $p^i(\ell_1,\dots,\ell_n,x_{s_1},\dots,x_{s_n}) = p^i(\ell_1,x_{s_1}) \circ \dots \circ p^i(\ell_n,x_{s_n})$ where $\circ$ indicates ``concatenation''. Without loss of generality we can consider just the special case of~$\ell_i=0$ and $\ell_i \in (0,\tau)$:
\begin{itemize}[topsep=0pt,itemsep=-1ex,partopsep=1ex,parsep=1ex,after=\vspace*{-\baselineskip}]
	\item \textbf{Transition}, for $\kappa \in \NN$:\\[-0.5cm]
	$$
	p^i(0,(\lambda_j,s_k),\kappa \tau) = (0,(\lambda_j,\kappa \tau))\,,
	$$
	\item \textbf{Vertex}, for $r\in (\kappa \tau, (\kappa+1) \tau)$:\\[-0.5cm]
	$$
	p^i(\ell_i,(s_k,r)) = (\ell_i,(v_\star, r)) = (\ell_i,r)\,.
	$$
\end{itemize}
Figure~\ref{fig:input-output-sheaf-maps} shows how a state sheaf is mapped, through the~$p^i$, to an input sheaf.

For the output map~$p^o$ we have a very similar situation. Following the same discussion as above, we need to focus only on the behavior of the output sheaf map~$p^o$ at transitions and vertices. We have:
\begin{itemize}[topsep=0pt,itemsep=-1ex,partopsep=1ex,parsep=1ex,after=\vspace*{-\baselineskip}]
	\item \textbf{Transition}, for $\kappa \in \NN\,, \omega_a = O(s_k) \wedge \omega_b = O(T(\lambda_j,s_k))$:\\[-0.5cm]
	$$
	p^o(0,(\lambda_j,s_k),\kappa \tau) = (0,h(\lambda_j,s_k),\kappa \tau) = (0,(\omega_a,\omega_b),\kappa \tau)\,,$$
	\item \textbf{Vertex}, for $r\in (\kappa \tau, (\kappa+1) \tau)$:\\[-0.5cm]
	$$p^o(\ell_i,(s_k,r)) = (\ell_i,(O(s_k),r)) = (\ell_i,(\omega_a,r))\,, $$
\end{itemize}
where~$h:\Lambda \times S \to \Omega \times \Omega$ was defined in~\eqref{eq:input_output_state_model}. An example on how~$p^o$ acts on a state sheaf is shown on the left of Figure~\ref{fig:input-output-sheaf-maps}. \\[-0.5cm]

\begin{remark}
As it might appear evident, after going through this construction, the modeling proceeded by first defining $\Int$-sheaves, each modeling in an independent fashion the input, output and state behaviors. We then ``align'' such sheaves by building input and output sheaf maps. The overall model is then an abstract machine. Although this construction can certainly appear very laborious, it is also very mechanical and suitable to be automated. Furthermore, one should notice the fact that this framework allows us to build machines starting from reusable building blocks. It would not be difficult, for example, to create a new machine with a different behavior but same input/output (of course we need to have the same number of states), by just changing the $\Int$-sheaf representing the state and adapting the input-output maps. Given that we did not constraint the output sequence to be the specific one produced by \ac{ACAS}---indeed, as we said, ~$\mathcal{K}(\Omega)$ is an arbitrary sequence of output symbols---we do not need to change it. The output sheaf map will ``take care'' of connecting the state sections with allowed output sections.
\end{remark}

\subsection{Aircraft Dynamics}

We are exclusively interested to model the vertical dynamics of a aircraft as it changes its altitude because of the change of the elevator deflection angle, obtained by having the pilot acting on the yoke.

Under simplifying assumptions, see Appendix~\ref{apx:aircraft} for the details, we have that the state vector consists of three valuables $(\alpha,q,\theta)^T$, angle of attack, pitch rate and thrust speed. As we are interested about the altitude of the aircraft, we can add another state variable~$h$ to the system of equations above with dynamics $\dot{h} = u \sin{\theta} \approx u \theta$, where we have made the assumption that $\theta$ is small which typically is reasonable for commercial aircrafts. Under these conditions, we can write the longitudinal dynamics compactly as the following linear system:
\begin{align*}
	\dot{\x} &= \A \x + \B \u\\
	\y &= \C \x
\end{align*}
where $\x = (\alpha,q,\theta,h)^T$ and $\y = h$. 

Thus we have that $\x \in S_\CDS \equiv \RR^4$, $\u \in I_\CDS \equiv\RR$ and $\y \in O_\CDS \equiv \RR$. Following Section~\ref{sec:interval_sheaves}, we can associate to~$I_\CDS$ and~$O_\CDS$ two $\Int$-sheaves, $\tilde{I}_\CDS$ and $\tilde{O}_\CDS$. In particular, for the dynamical system in consideration we have that the sections are, for the input, $\tilde{I}_\CDS(\ell) = \{c:[0,\ell] \to I_\CDS| c \in C^\infty\}$,
and similarly for the output $\tilde{O}_\CDS(\ell) = \{c:[0,\ell] \to O_\CDS| c \in C^\infty\}$.

Thus we have that the airplane dynamics are modeled as the tuple $\mathcal{A}=(S, f^\dyn, f^\rdt)$ where $f^\dyn: I_\CDS \times S_\CDS \to S_\CDS$ is the linear map described by the pair $(\A\in \RR^{4 \times 4}, \B \in \RR^{4})$, and $f^\rdt:S_\CDS \to O_\CDS$ is the linear map described by $\C \in \RR^{1\times 4}$.

We can then define a sheaf $\tilde{S}_\CDS \in \IntSh$ whose $\ell$-length sections are 
\begin{align*}
	\tilde{S}_\CDS(\ell) = \{&(\u,\x):[0,\ell] \to I_\CDS \times S_\CDS | \dot{x} = \A \x + \B \u,\\
	& \x \in C^\infty, \u \in C^\infty\}\,.
\end{align*}

We can thus model the aircraft as the following abstract machine defined by the span:
$$
\begin{tikzcd}[row sep = 0.3cm]
\tilde{I}_\CDS \in \IntSh& \tilde{S}_\CDS \in \IntSh \ar[l, "\pi_1"'] \ar[r,"f^\rdt \circ \pi_2"]& \tilde{O}_\CDS \in \IntSh\,.
\end{tikzcd}
$$
%

\subsection{Composition}

We are interested now to compute the composition of the two machines: 1) the \ac{ACAS} logic and 2) the aircraft dynamics. We underline once more that the two machine have very different models of computations, and the have been both abstracted into abstract machines.

Before doing this we need to define a new machine---the ``human''---that translates the output sheaf~$\tilde{O}_\LTS$, into a new sheaf~$\tilde{\Delta}$ in which $\ell$-length sections have values in $\{-\bar{\delta},0, +\bar\delta\}$ where $\pm\bar\delta$ are the deflector position corresponding to descent, level flight and climb. 

This machine is obviously very simple and readily defined by the following commutative diagram:
$$
\begin{tikzcd}
	\tilde{O}_\LTS & \tilde{O}_\LTS \ar[l,equal] \ar[r,"\phi"] & \tilde{\Delta}
\end{tikzcd}
$$
where $\phi((\ell_1,\dots,\ell_n,x_{o_1},\dots,x_{o_n})) = (\ell_1,\dots,\ell_n, x_{\delta_1},\dots,\allowbreak x_{\delta_n})$ and where we have that
\begin{align*}
	\phi(\ell_i,x_{o_i}) =
	\begin{cases}	
		\phi(\ell_i,((0,((\omega_a ,\omega_b),\kappa \tau))) & \text{$\kappa \in \NN$}\,,\\
		\phi(\ell_i,((0,(\ell_i,(\omega_a,r))) &\text{$r \in (0,\tau)$}\,,
	\end{cases}
\end{align*}
where
\begin{align*}
 &\phi(\ell_i,((0,((\omega_a ,\omega_b),\kappa \tau))) =\\
 &\begin{cases}
 (0,((0,0),\kappa \tau)) & \text{if $\omega_a = \omega_b = \omega_1$,}\\
 (0,((0,+\bar\delta),\kappa \tau)) & \text{if $\omega_a = \omega_1$, $\omega_b = \omega_2$,}\\
 (0,((0,-\bar\delta),\kappa \tau)) & \text{if $\omega_a = \omega_1$, $\omega_b = \omega_3$,}\\
 \dots & \dots \\
 (0,((-\bar\delta,0),\kappa \tau)) & \text{if $\omega_a = \omega_3$, $\omega_b = \omega_1$,}
 \end{cases} \\
 \intertext{and}
 &\phi(\ell_i,((0,(\ell_i,(\omega_a,r))) =
 \begin{cases}
 (\ell_i,(0,r)) & \text{if $\omega_a = \omega_1$,}\\
 (\ell_i,(+\bar\delta,r)) & \text{if $\omega_a = \omega_2$,}\\
 (\ell_i,(-\bar\delta,r)) & \text{if $\omega_a = \omega_3$.}
 \end{cases} 
\end{align*}
Thus, a section of the output of such machine, just ``maps'' the labels $\omega_i$ to the set $\{0,\pm \delta\}$.

We can then compose these two machines into a new machine via pullback, as we discussed in Section~\ref{subsec:composition}. In particular, we obtain
$$
\begin{tikzcd}[column sep=0.01cm,row sep=0.1cm]
    && \tilde{S}_\LTS \times_{\tilde{O}_\LTS} \tilde{O}_\LTS \ar[ldd,"\pi_1"'] \ar[rdd,"\pi_2"] \ar[dddd,phantom,"\lrcorner" very near start] && && &\tilde{S}_\LTS \ar[ldddd,"p^i"'] \ar[rdddd,"p^o \circ \phi"] &\\\\
  	& \tilde{S}_\LTS \ar[ldd, "p^i"'] \ar[rdd,"p^o"] & & \tilde{O}_\LTS \ar[ldd,equal] \ar[rdd,"\phi"] & &\hspace*{-0.4cm}\simeq\hspace*{-0.4cm} & & & \\\\
  	  \tilde{I}_\LTS & & \tilde{O}_\LTS	& & \tilde{\Delta} && \tilde{I}_\LTS	& & \tilde{\Delta}
\end{tikzcd}
$$ 
where the isomorphism holds because of the special structure of the machine modeling the human. 

We now need to compose the above machine, which is itself the composition of the machine modeling the \ac{ACAS} logic and the human, with the continuous dynamical system, representing the aircraft dynamics.

In order to be able to do this we need to consider a slightly extended version of the continuous dynamical system that can have piecewise constant inputs. Given the lack of space and the fact that this extension is fairly obvious, we refer the reader to~\cite[End of Section 5.1]{DS-CV-PS:16}. This allows us to build the following machine
$$
\begin{tikzcd}
	\tilde{\Delta}& \tilde{\Delta} \ar[l,equal] \ar[r, "g"]&\tilde{I}_\CDS
\end{tikzcd}
$$
which allows piecewise constant elevator deflection angles to be inputs to the continuous dynamical system (with abuse of notation~$\tilde{I}_\CDS$ from here onwards indicates an $\Int$-sheaf whose sections are piecewise constant signals).

For the full composition we are then considering the following commutative diagram:
$$
\begin{tikzcd}[column sep=0.2cm,row sep=0.01cm]
 &&& \tilde{S}_\CMP \ar[rrddd,"\pi_2"] \ar[llddd,"\pi_1"'] \ar[dddd,phantom,"\lrcorner" very near start] &&&\\\\\\
 & \tilde{S}_\LTS \ar[ldd,"p^i"'] \ar[rdd,"p^o \circ \phi"]&&&& \tilde{S}_\CDS \ar[ldd,"\pi_1"'] \ar[rdd,"f^\rdt \circ \pi_2"]&\\
 &&& \tilde{\Delta} \ar[ld,equal] \ar[rd, "g"] &&&\\
 \tilde{I}_\LTS && \tilde{\Delta} && \tilde{I}_\CDS && \tilde{O}_\CDS 
\end{tikzcd}
$$
where $\tilde{S}_\CMP = \tilde{S}_\LTS \times_{\tilde{I}_\CDS} \tilde{S}_\CDS$. Thus we have that a section of the composed system is given by
\begin{align*}
	&\tilde{S}_\CMP(\ell) = \{(s_\LTS,s_\CDS) \in \tilde{S}_\LTS(\ell) \times \tilde{S}_\CDS(\ell)|\\
	& (p^o \circ \phi \circ g)(s_\LTS) = s_\CDS\}\,.
\end{align*}
\begin{figure}[t]
	\centering
	\ifdefined \doublecol
		\includegraphics[width=0.8\hsize]{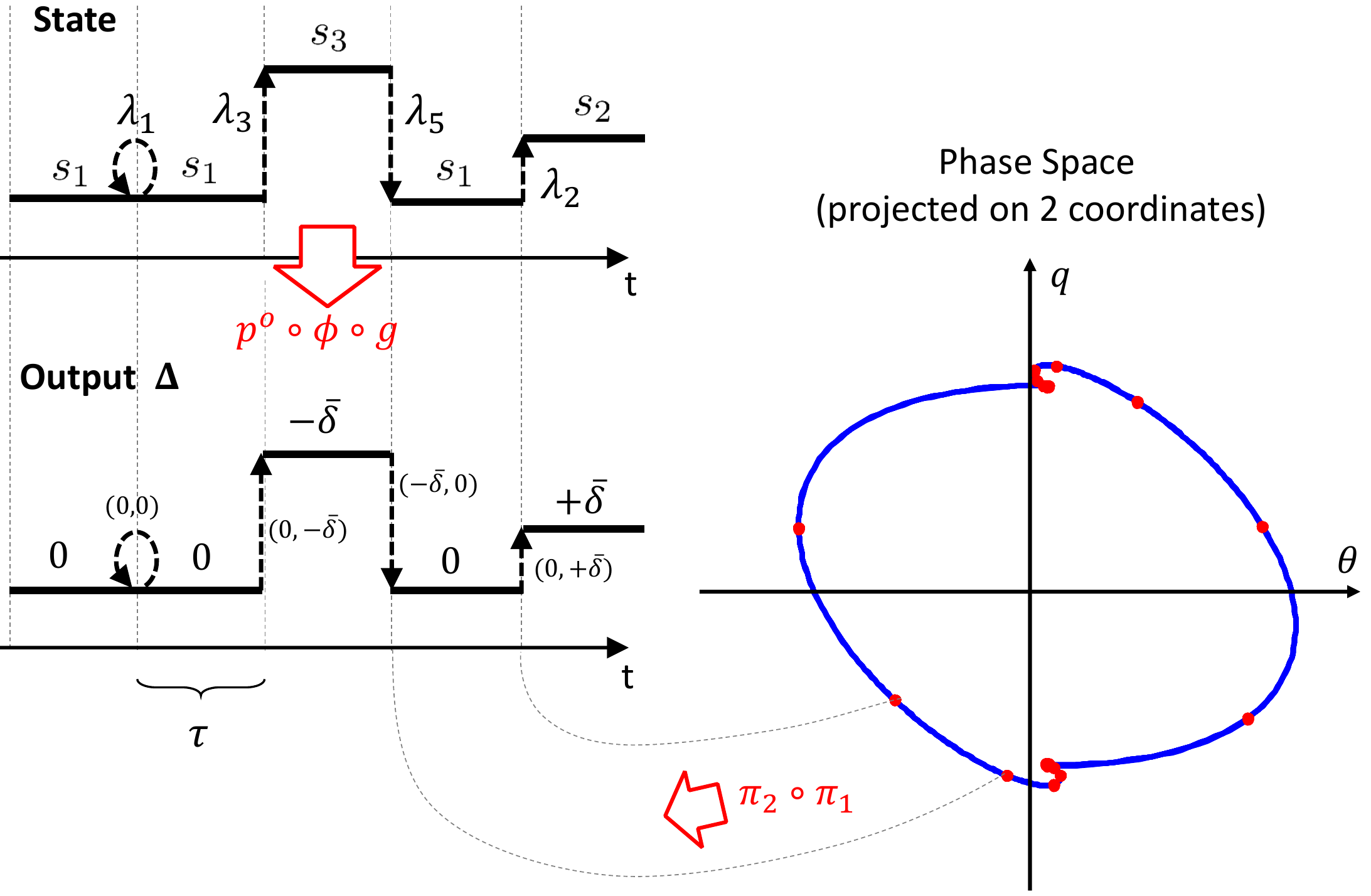}
	\else
		\includegraphics[width=0.75\hsize]{composition.pdf}
	\fi
	\caption{Pictorial representation of the state of the composed machine $\tilde{S}_\CMP$. The phase space corresponds to a ``square wave'' (climb, level, descend, climb, etc.) used to exemplify the connection between the labeled transition system and the continuous dynamics.}\label{fig:state_composed}
\end{figure}
Figure~\ref{fig:state_composed} depicts, at a high level, what is happening once the machines are combined. In particular it shows how the internal states of the ``discrete'' and ``continuous'' parts maps to the same input. Clearly, with no surprise the overall machine is a hybrid system, in the same sense of~\cite{Henzinger:96}. However, it is interesting to see that we did not define the hybrid automaton in advance --- deciding what is the discrete state, what are the dynamics for each state, etc. --- it all emerged the proposed abstraction and composition framework.

For this simple example, the approach appears overwhelmingly complicated, with little or no gain. We stress once more, that the power of this approach is that once subsystems are abstracted into abstract machines, the designer can ``mix-and-match'' subcomponents, as long as one can find sheaf maps comparing the output sheaves and input sheaves, see Figure~\ref{fig:overall_idea}. The user does not need to select a common abstract model a priori, i.e.\ he/she does not need to decide whether the best common model is a labeled transition system or a hybrid system or a Petri Net, etc. All systems are all mapped to a consistent and composable representation, reliving the design of ad-hoc choices. 

\section{Contracts}
\label{sec:contracts}
So far we have discussed how we can leverage the proposed framework to abstract and composed various models of computations representing subsystems in a SoS. Of course, although it provides a solid formalism, unless it also enables analysis, its relevancy would be fairly limited. 

It turns out that the category of sheaves we have considered so far, is more precisely a (Grothendieck) topos, namely a nice category that has $\mathbf{Set}$-like properties. As in~$\mathbf{Set}$ one can define a characteristic functions $\chi_A:X \to \{0,1\}$, for $A \subseteq X \in \mathbf{Set}$, in toposes there is an object, called \emph{subobject classifier}, that enables us to classify objects accordingly to given properties. Specifically for the behavior type~$\IntSh$, there is a behavior type~\texttt{Prop} that classifies sections based on a given property. Every topos has an associated internal language and higher-order logic. It supports the standard logical connectives $\top, \bot, \wedge, \vee, \neg, \Rightarrow, \Leftrightarrow$ and quantifiers, $\forall, \exists$. The logic (called Heyting's calculus) however, is intuitionistic/constructive so the law of excluded middle ($P \wedge \neg P$) and the double negation elimination ($\neg \neg P$) do not hold in general. 

Note that given the definition of a behavior, namely a $\Int$-sheaf, properties/contracts will represent \emph{safety properties}: if a system satisfies a given property over an interval of time, it must satisfy it for every sub-interval (recall that an $\Int$-sheaf describes behavior over an interval and \emph{every} subinterval of the given interval).

With this in mind, behaviors (sheaves) become types in the higher-order logic and predicates can be proved using a set of axioms. The expressiveness of the logic with semantics in $\Int$-sheaves enables us to also define time derivatives.

In~\cite{schultz-spivak:2017} a set of axioms have been developed from which one can prove more complex propositions. The Joyal-Kripke sheaf semantics applies, although it simplifies in the context of $\Int$-sheaves. In~\cite[Chapter 5]{schultz-spivak:2017} this is made explicit.

From a more practical perspective, one can ``neglect'' the fact that sheaves are types in the logic, and reason in a more ``standard'' fashion. For example, for the airplane subsystem, we can express the contract ``given commands ``climb'', ''level'' or ''descend'' the pitch angle rate changes of a certain amount, ``rate'''' as:
\begin{align*}
  (\texttt{P:Cmnd})(\theta:\mathds{R})&(\texttt{rate}:\mathds{R}) \vdash (\texttt{P = level} \Rightarrow \dot{\theta} = 0)\\ 
  &\wedge (\texttt{P = descend} \Rightarrow \dot{\theta} = -\texttt{rate}) \\
  &\wedge (\texttt{P = climb} \Rightarrow \dot{\theta} = +\texttt{rate})\,.
\end{align*}

Note that, for example $\texttt{P:Cmnd}$ needs to be interpreted as an $\Int_N$-sheaf, namely the output of the abstract machine modeling the \ac{ACAS} logic. Thus $\texttt{Cmnd}$ represent paths (sequence of commands) on the transition system~$\mathcal{T}$. Saying that $\texttt{P = descend}$ implies that if the $\Int_N$-sheaf $P$ is the constant sheaf $\texttt{descend}$, then the decent rate is $\texttt{-rate}$. More specifically, if the behavior is $\texttt{descend}$, within a certain time interval then the pitch rate is negative and with that we mean that for any subinterval of time the command is $\texttt{descend}$ and the descend rate $\texttt{-rate}$.

One important point to note is that time is built into logic (through interval sheaves), meaning that propositions can contain time explicitly. We believe, but this has not been proved yet, that Metric Temporal Logic (MTL) and derived logics can be embedded in the proposed logic. It has been shown to hold true for linear temporal logic (LTL)~\cite{schultz-spivak:2017}, thus one can take advantage of the~$\Int$-sheaf formalism, while retaining decidability.

The last point we want to stress here is that, as types in the logic are behaviors over intervals, we do not need to discretize continuous dynamics to be able to prove properties. However, the cost for this is the loss of decidability. However, recent theorem provers software packages~\cite{lean:2015} can aid a designer verifying properties. 

\section{Conclusions}
\label{sec:conclusions}

This paper introduces a new framework based on interval sheaves to describe the behavior of systems. Behaviors can then be ``linked'' together to form a very general abstraction, abstract machines, that can be formally composed accordingly to an interconnection (wiring) diagram. The proposed framework shifts the problem of abstraction upfront, where each subsystem is first abstracted, providing the benefit of making composition less of an ad-hoc design process.

To ground the discussion we have shown how this could be applied to continuous and labeled transition systems using a simple~\ac{ACAS} example. While at this stage no software exists to automate this abstraction, we believe that some of the required procedures can be automated.

We concluded the paper showing that the deep connection between sheaf and topos theory offers us a way to define a higher-order temporal logic that will be instrumental to analyze systems and represent properties, contracts and requirements. 

\bibliographystyle{plain}
\bibliography{refs} 

\newpage

\appendix

\section{Graphs and Reflexive Graphs}
\label{apx:graphs}
In the paper we will consider graphs defined as $\mathcal{G} = (V,E,\allowbreak \src,\tgt)\in \cat{Grph}$ where~$V$ and $E$ are sets (of ``vertices" and ``edges"). The source, $\src\colon E\to V$ and target~$\tgt\colon E\to V$ functions serve to assign each $e \in E$ an ordered pair of vertices. We will often denote a graph as
$$
\mathcal{G} = (\doublearrow{E}{V})\,.
$$
Another category of graphs that will play an important role is that of \emph{reflexive graphs},~$\rGrph$. These are defined as
$$
\mathcal{G} = (\doublearrowids{E}{V})\,,
$$
namely, each vertex $v$ has a designated self-loop $\ids(v)\in E$.

\section{Aircraft Collision Avoidance System}
\label{apx:acas}
An \ac{ACAS} system installed on an aircraft uses information received from aircrafts in its vicinity to detect violations of \emph{safe separation}. If it is the case that a collision may occur, the \ac{ACAS} will provide the pilot with an advisory, called \ac{TA}. Generally a \ac{TA} will not require a pilot to change the course, but requires the pilot to be prepare to take an action. The \ac{ACAS} system will be estimating the collision time (based on relative speeds and altitudes) and when this is below a certain threshold, a \ac{RA} will be issued. This requires the pilots to take an action to avoid the intruder. In the cooperative case, the \ac{ACAS} will suggest a vehicle to climb and the other to descend\footnote{In the uncooperative setting the TCAS makes the assumption that the intruder aircraft maintains the same altitude.}. Furthermore, \ac{ACAS}, has built in a \emph{reversal} function that reverses the \ac{RA} decision the intruder aircraft does not comply with the initial \ac{RA}.

\begin{figure}[h]
	\centering
	\ifdefined \doublecol
		\includegraphics[width=0.8\hsize]{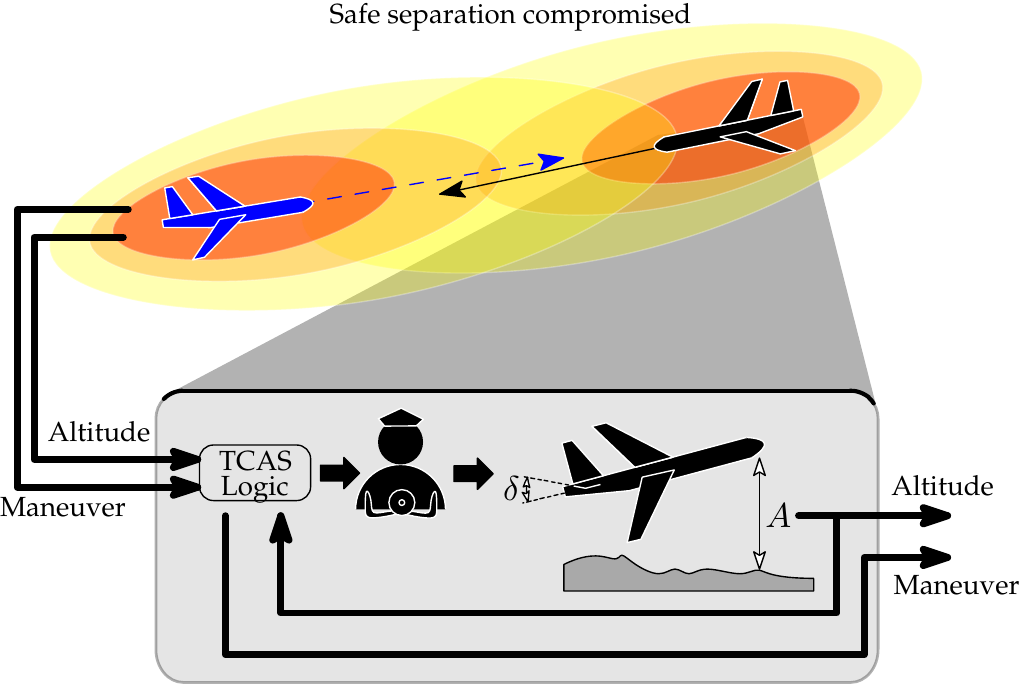}
	\else
		\includegraphics[width=0.60\hsize]{ACAS_system.pdf}
	\fi
	\caption{High level view of \ac{ACAS} and main subsystems within one system (aircraft).}\label{fig:ACAS_example}
\end{figure}

Figure~\ref{fig:ACAS_example}, shows more in detail the problem we are considering 

\section{Modeling the state as~$\IntSh$}
\label{apx:state}

Given $G_S= (\Lambda \times S \rightrightarrows S)$ we construct the $\Int$-sheaf $V_S = S \times \Yon_\tau$, where we remind that $S=\{s_1,s_2,s_3\}$ and $\Lambda=\{\lambda_1,\dots,\lambda_5\}$. We also construct the set $E_S = (\Lambda \times S) \sqcup V_S(0)$ where $V_S(0) = S \times \Yon_\tau(0)$. We then define the following reflexive graph: 
$$\vspace*{-0.cm}
\begin{tikzcd}[column sep=0.05cm, row sep = 0.1cm]
G'_S:\hspace*{-0.2cm} &\Lambda \times S 
\ar[rrrr, "{\src'(\lambda_j,s_i)=(\src(\lambda_j,s_i),\tau)=(s_i,\tau)}" description,  bend left=55] 
\ar[rrrr, "{\tgt'(\lambda_j,s_i)=(\tgt(\lambda_j,s_i),\tau)=(T(\lambda_j,s_i),\tau)}" description, bend left=20] & \hspace*{-0.2cm}\bigsqcup\hspace*{-0.2cm}  & S \times \Yon_\tau(0) 
\ar[rr,"{\src'(s_i,r)=(s_i,r)}"'  description, bend right=45],  
\ar[rr,"{\tgt'(s_i,r)=(s_i,r)}" description, bend right=20] &\hspace*{2cm}& S \times \Yon_\tau(0)  
\ar[ll,"{\ids(s_i,r)=(s_i,r)}"'  description]
\end{tikzcd}
$$
where $0 \leq r \leq \tau$ and $T(\lambda_i,s_j)$ is the state we transition to starting from~$s_j$ when the input is~$\lambda_i$. Recall that $T : \Lambda \times S \to S$ is the state transition function of the labeled transition system~\eqref{eq:input_output_state_model}.

The hybrid sheaf datum associated is then $H_S=(V_S,G'_S)\in\rgInt$. We can derive the $\Int$-presheaf~$\bar{H}_S$ whose $\ell$-length sections are given by the pullback in~$\Set$, $\bar{H}_S(\ell) = V_S(\ell)\allowbreak \times_{V_S(0) \times V_S(0)}\allowbreak E_S \times E_S$. Given that~$\Yon_\tau(\ell)=\emptyset$ for $\ell > \tau$ any $\ell$-length section is defined of having length of either $\tau$-length or more.

The restriction maps are, for $r \in [p,\tau-q]$ and let $s_k = T(\lambda_j,s_i)$, then
\begin{align*}
&\bar{H}_O(\Trp)(((\lambda_j,s_i),0),(s_k,r),((\lambda_h,s_k),\tau)) = \\
&\begin{cases}
(\ids(s_k,p),(s_k,r),((\lambda_h,s_k),\tau)) & \text{if $p \neq 0$ and $q = 0$,}\\
((\lambda_j,s_i),0),(s_k,r),\ids((s_k,q))) & \text{if $p =0$ and $q \neq 0$,}\\
(\ids(s_k,p),(s_k,r), \ids(s_k,q)) & \text{if $p\neq0$ and $q\neq0$.}
\end{cases}
\end{align*}
The sheafification of the presheaf~$\bar{H}_S$ into a $\Int$-sheaf by the representative functor~$R$ proceeds as in the previous two cases to obtain the $\Int$-sheaf $\tilde{S}$, whose sections are piecewise constant signals whose transitions are triggered by~$\lambda_i$ and the constant values are associated to the state~$s_i$.

\section{Longitudinal Dynamics of an Aircraft}
\label{apx:aircraft}
Following~\cite{Stengel:15}, under small perturbation assumption we can decouple the lateral and longitudinal dynamics. Given that \ac{ACAS} is only providing the pilot with an avoidance action on the longitudinal plane, we just need to consider such dynamics.

\begin{figure}[h]
	\centering
	\ifdefined \doublecol
	\includegraphics[width=1\hsize]{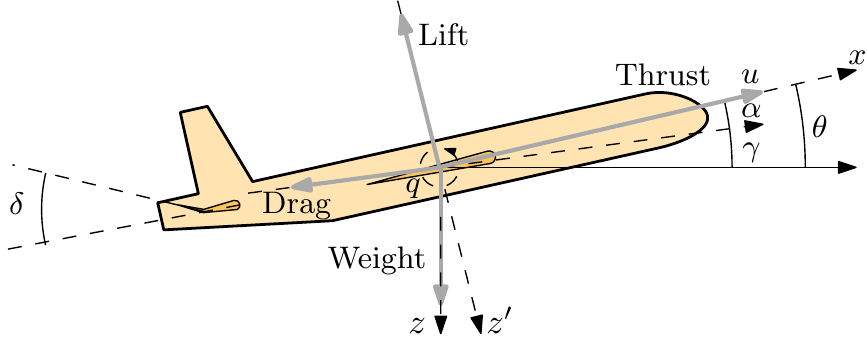}
	\else
	\includegraphics[width=0.6\hsize]{aircraft.pdf}
	\fi
	\caption{Main axis describing the longitudinal dynamics.}\label{fig:airplane_pitch}
\end{figure}

With reference to Figure~\ref{fig:airplane_pitch}, we assume that the aircraft is in steady-cruise at constant altitude and velocity. In this setting the thrust, drag, weight and lift forces balance each other in the~$x$ and~$z$ directions, respectively.

We also make the simplifying assumption that a change in pitch angle does not change the aircraft speed.

Under such simplifying assumption, we have that the equation of motion are:
\begin{align*}
\dot{u} &= X_u u + X_\alpha - X_0 \alpha_0 q -g \cos\theta_0 \theta + X_{\delta_t} \delta_t \\
\dot{\alpha} &= \frac{Z_u}{U_0} u + \frac{Z_\alpha}{U_0} \alpha + \frac{U_0+Z_q}{U_0} q -g \frac{\sin{\theta_0}}{U_0}  \theta + \frac{Z_{\delta_e}}{U_0} \delta_e + \frac{Z_{\delta_t}}{U_0} \delta_t\\
\dot{q} &= M_u u + M_\alpha \alpha + M_q q + M_{\delta_e} \delta_e + M_{\delta_r} \delta_t\\
\dot{\theta} & = q
\end{align*}
where we have~$u$ begin the thrust speed,~$\alpha$ the angle of attack,~$q$ the pitch rate and~$\theta$ the pitch angle defection, respectively.

We have that $X_\bullet$,~$Z_\bullet$ and~$M_\bullet$ are the longitudinal stability derivatives with respect to the corresponding state variables. The control inputs are~$\delta_e$ and~$\delta_t$ representing the deflection of the elevator and the thrust maneuvering. Further we have~$U_0$ and~$\theta_0$ be the wind speed and pitch angle of a trimmed state, and $g$ is gravity.

Here we make further simplifying assumptions: 1) the thrust~$\delta_t$ is constant and equal to~$u$ and (2) we neglect that control actions change the vehicle speed~$u$.

\end{document}